\documentclass{article}
\usepackage{amssymb}
\usepackage{color}

\newcommand{\II}{\mbox{\scriptsize\sc II}}

\newcommand{\IV}{\mbox{\scriptsize\sc IV}}
\newcommand{\V}{\mbox{\scriptsize\sc V}}
\newcommand{\VI}{\mbox{\scriptsize\sc VI}}

\newcommand{\comment}[1]{}
\newcommand{\eqn}[2]{\[\begin{array}{#1}#2\end{array}\]}

\newcommand{\dfrac}[2]{{\displaystyle\frac{#1}{#2}}}
\newcommand{\br}[1]{\langle #1 \rangle}
\newcommand{\pbr}[1]{\{ #1 \}}
\newcommand{\pad}{\mbox{ad}_{\{\}}}

\newcommand{\BZ}{{\mathbb Z}}
\newcommand{\BC}{{\mathbb C}}
\newcommand{\CK}{{\mathcal K}}
\newcommand{\CA}{{\mathcal A}}
\newcommand{\bd}[1]{\mbox{\boldmath $#1$}}
\newcommand{\dsum}[1]{{\displaystyle \sum#1}}
\newcommand{\dprod}[1]{{\displaystyle \prod#1}}
\newtheorem{thm}{Theorem}[section]
\newtheorem{prop}[thm]{Proposition}
\newtheorem{lem}[thm]{Lemma}
\newtheorem{cor}[thm]{Corollary}

\def\be{\begin{equation}}
\def\ee{\end{equation}}
\def\proof{\noindent{\it Proof.}\hskip2mm}
\def\qed{$[]$}

\title{$q$-Painlev\'e systems arising from $q$-KP hierarchy}
\author{
Kenji KAJIWARA${}^{1)}$, 
Masatoshi NOUMI${}^{2)}$ and Yasuhiko YAMADA${^{2)}}$\\[5mm]
{\small ${}^{1)}$ Graduate School of Mathematics, Kyushu University,}\\
{\small 6-10-1 Hakozaki, Higashi-ku, Fukuoka 812-8512, Japan}\\[3mm]
{\small  ${}^{2)}$ Department of Mathematics, Kobe University}\\
{\small 1-1 Rokko, Kobe 657-8501, Japan}}
\begin{document}
\maketitle
\begin{abstract}
A system of $q$-Painlev\'e type equations with multi-time variables
$t_1,\ldots,t_M$ is obtained as a similarity reduction of the
$N$-reduced $q$-KP hierarchy.  This system has affine Weyl group
symmetry of type $A^{(1)}_{M-1} \times A^{(1)}_{N-1}$.  Its rational
solutions are constructed in terms of $q$-Schur functions.
\end{abstract}

\section{Introduction}

Fix an integer $N\ge 3$.  The following system of $q$-difference
equations has been introduced in \cite{KNY2}:
\begin{equation}\label{qPAn}
\begin{array}l
\overline{a_i}=a_i, \quad
\overline{x_i}=\dfrac{x_{i-1}}{a_i}\dfrac{\varphi_{i-1}(x)}
{\varphi_{i+1}(x)}, \\[4mm]
\varphi_i(x)=1+x_{i-1}+x_{i-2}x_{i-1}+\cdots+ x_{i-N+1}\cdots x_{i-1},
\end{array}
\end{equation}
for $i=0,1,\ldots,N-1$.
Here $a_i$ and $x_i$ ($i=0,1,\ldots,N-1$) are constant parameters and
dependent variables satisfying $a_{i+N}=a_i$ and $x_{i+N}=x_i$,
respectively.  The symbol $\overline{f}$ denotes the discrete time
evolution of a variable $f$.  We put $a_0a_1\cdots a_{N-1}=q^{-N}$ and
$x_0x_1\cdots x_{N-1}=t$; $t$ plays the role of independent variable
such that $\overline t=q^N t$.

The system (\ref{qPAn}) has the B\"acklund transformations $s_0,s_1,\ldots,s_{N-1},\pi$
defined as follows:
\begin{equation}
\begin{array}{ll}
\pi(a_i)=a_{i+1}, &\pi(x_i)=x_{i+1}, \\
s_i(a_{i-1})=a_{i-1}a_i,\quad &
s_i(x_{i-1})=\dfrac{x_{i-1}}{a_i}\dfrac{1+a_i x_i}{1+x_i},
\\
s_i(a_i)=\dfrac{1}{a_i},&
s_i(x_i)=a_i x_i,
\\
\smallskip
s_{i}(a_{i+1})=a_i a_{i+1},&
s_i(x_{i+1})=x_{i+1}\dfrac{1+x_i}{1+a_i x_i},
\\
s_i(a_j)=a_j, & s_i(x_j)=x_j \qquad(j\ne i,i\pm1).
\end{array}
\end{equation}
These transformations generate the affine Weyl group of type $A^{(1)}_{N-1}$.

The case of $N=3$ corresponds to the $q$-Painlev\'e IV equation ($q$-$P_{\IV}$)\cite{KNY}
through the change of variables $a_i\to a_i^2$, $x_i \to {f_i}/{a_i}$ and $q\to q^{-1/3}$.
The case of $N=4$ is considered as a $q$-difference analog of Painlev\'e
V equation ($q$-$P_{\V}$), whose rational solutions are studied in \cite{Mas}.

The aim of this paper is to introduce a hierarchy of these equations
with multi-time variables $t_1,\ldots,t_M$.  This hierarchy, formulated
as a similarity reduction of an $N$-reduced $q$-KP hierarchy, has the
affine Weyl group symmetry of type $A^{(1)}_{M-1} \times A^{(1)}_{N-1}$.
We also construct certain rational solutions expressed in terms of $q$-Schur
functions by using this formulation.

\section{$q$-KP hierarchy}

We first introduce a $q$-difference analog of the KP hierarchy ($q$-KP
hierarchy). 

Let $t_i$ and $T_i$ ($i=1,2,\ldots, M$) be the time variables and
their $q$-shift operators: 
\begin{equation}
T_i(t_i)=q t_i, \quad T_i(t_j)=t_j \quad (i \neq j).
\end{equation}
We also use the notation such as $T_iT_j\cdots T_k=T_{i,j,\ldots,k}$.
For $i=1,2,\ldots,M$, we define the $\BZ \times \BZ$ matrices
\begin{equation}
B_i=U_i+t_i \Lambda,
\end{equation}
where
\begin{equation}
U_i={\rm diag}(u_{i,j})_{j \in \BZ}, \quad
\Lambda=(\delta_{i+1,j})_{i,j \in \BZ},
\end{equation}
and consider the linear $q$-difference equations
\begin{equation}
T_k \Psi=B_k \Psi, \quad (k=1,\ldots,M) \quad \Psi=(\psi_i)_{i \in \BZ}.\label{qKPLax}
\end{equation}
We define the $q$-KP hierarchy as a system of nonlinear $q$-difference
equations for the unknown variables $u_{i,j}=u_{i,j}(t_1,\ldots,t_M)$,
through the compatibility condition of (\ref{qKPLax}),
\begin{equation}
 T_k(B_i)B_k = T_i(B_k)B_i\qquad (i,k=1,\ldots,M). \label{qKPZS}
\end{equation}
The evolution equation for $u_{i,j}$ with respect to $t_k$ is expressed explicitly as
\begin{equation}\label{qKPu}
T_k(u_{i,j})=u_{i,j}\dfrac{t_i u_{k,j+1}-t_k u_{i,j+1}}
{t_i u_{k,j}-t_k u_{i,j}} \quad (i \neq k).
\end{equation}
Introduce the $\tau$-functions $\tau_j$ ($j\in \BZ$) by
\begin{equation}
u_{i,j}=\dfrac{T_i(\tau_j)\tau_{j-1}}{T_i(\tau_{j-1})\tau_j}.\label{qKPtau}
\end{equation}
These $\tau$ functions are defined up to normalization constants.
By substituting (\ref{qKPtau}) into (\ref{qKPu}), we have
\begin{thm}
Under an appropriate normalization, the $q$-KP hierarchy
(\ref{qKPu}) is described by the Hirota-Miwa bilinear $q$-difference equations
\begin{equation}
\label{H-M}
t_iT_i(\tau_{k-1})T_j(\tau_k)-t_j T_j(\tau_{k-1})T_i(\tau_k)=
(t_i-t_j)T_{ij}(\tau_{k-1})\tau_k.
\end{equation}
\end{thm}

A class of rational solutions of the $q$-KP hierarchy is described
in terms of the $q$-Schur functions as follows.
Define $h_k(t)$ ($k=0,1,2,\ldots $) by
\begin{equation}
\dsum{_{k=0}^\infty} h_{k}(t)z^k
=\dprod{_{i=1}^M} \dfrac{1}{(t_i z;q)_{\infty}},
\end{equation}
or
\begin{equation}
h_{k}(t)=\dsum{_{\nu_1+\cdots+\nu_M=k}}
\dfrac{t_1^{\nu_1}\cdots t_{M}^{\nu_M}}{(q;q)_{\nu_1}\cdots(q;q)_{\nu_{M}}}.
\end{equation}
We set $h_k(t)=0$ $(k<0)$. For a sequence of integers
$\lambda=(\lambda_1, \lambda_2, \ldots,\lambda_l)$,
the $q$-Schur function $S_{\lambda}(t)$ is defined as
\begin{equation}\label{qSchur}
S_{\lambda}(t)
=\det\Big[ h_{\lambda_{i}-i+j}(t) \Big]_{1 \leq i,j \leq l}.
\end{equation}
For $\lambda=(\lambda_1,\ldots,\lambda_l)$ and $k \in \BZ$, we put
$(k,\lambda)=(k,\lambda_1,\ldots,\lambda_l)$.  For two sequences of
integers $\lambda=(\lambda_1,\ldots,\lambda_l)$ and
$\mu=(\mu_1,\ldots,\mu_m)$, we say that $\lambda\equiv \mu$ if
$S_\lambda=\pm S_\mu$. For instance we have
$(\ldots,a,b,\ldots)\equiv (\ldots,b-1,a+1,\ldots)$ and
$(\ldots,a,b,0)\equiv (\ldots,a,b)$.

The following Proposition asserts that the $q$-Schur functions solve the
Hirota-Miwa equation (\ref{H-M}).
\begin{prop}\label{qschur-sol}
For any $k$ and $\lambda$, we have
\begin{equation}\label{qKP-S}
t_iT_i(S_{\lambda}) T_j(S_{(k,\lambda)})-
t_jT_j(S_{\lambda}) T_i(S_{(k,\lambda)})
=(t_i-t_j) T_{ij}(S_{\lambda}) S_{(k,\lambda)}.
\end{equation}
\end{prop}

\proof 
Define a semi-infinite matrix $\Phi$ with indices $i,j=1,2,\ldots$ as
\begin{equation}
\Phi=\sum_{l=0}^{\infty} h_l(t) \Lambda^l,\quad
\Lambda_{i,j}=\delta_{i+1,j}.
\end{equation}
For any partition
 $\lambda=(\lambda_1,\lambda_2,\ldots,\lambda_l)$, $\lambda_1 \geq
 \lambda_2 \geq \ldots \geq \lambda_l\geq 0$,
the $q$-Schur function $S_{\lambda}(t)$ is the minor determinant of 
the matrix $\Phi$ with rows $(1,2,\ldots,l)$ and
columns $(\lambda_l+1,\lambda_{l-1}+2,\ldots,\lambda_1+l)$.
Note that the matrix $\Phi$ solves the $q$-difference system
\begin{equation}
T_n(\Phi)=(1-t_n\Lambda)\Phi\quad(n=1,2,\ldots,M).
\end{equation}
Then the relation (\ref{qKP-S}) reduces to the determinant identity
(Pl\"ucker relation) in Lemma \ref{matid} below for $X=\Phi \Sigma$
defined with an appropriate permutation matrix $\Sigma$.  \qed

\begin{lem}\label{matid}
For any matrix $X$, we have
\begin{equation}
(a-b){\det}_{n}(V_a V_b X)
{\det}_{n+1}(V_c X)+(\mbox{$abc$ \ {\rm cyclic}})=0.
\end{equation}
Where $V_a=1-a \Lambda$ and
${\det}_n(X)$ is the $n \times n$ minor determinant of the matrix
$X$ with rows and columns $(1,2,\ldots,n)$.
\end{lem}

We note that the $q$-Schur functions also satisfy the following bilinear
$q$-difference equation.
\begin{prop}\label{qS-Back}
For any partition $\lambda$ and $k<l \in \BZ$, we have
\begin{equation}
t_i S_{(l-1,k,\lambda)}T_i(S_{\lambda})
+S_{(k,\lambda)} T_i(S_{(l,\lambda)})
-S_{(l,\lambda)} T_i(S_{(k,\lambda)})=0.
\end{equation}
\end{prop}
\proof This is also a consequence of the following determinant identity.\qed
\begin{lem}
For any matrix $X$, we have
\begin{equation}
a \xi^{(I, k,l)}(X) \xi^I(V_a X)+
\xi^{(I,k)}(X)\xi^{(I,l)}(V_a X)-
\xi^{(I,l)}(X)\xi^{(I,k)}(V_a X)=0,
\end{equation}
where $\xi^J(X)$ is a minor determinant of $X$ with
rows $(1,2,\ldots,p)$ and columns $J=(j_1,\ldots,j_p)$.
\end{lem}

\section{Reduction to $q$-Painlev\'e equations}

We consider two kinds of reduction conditions for the $q$-KP hierarchy,
namely, the $N$-reduction and the similarity reduction.

We first impose $N$-periodicity,
\begin{equation}
u_{i,j+N}=u_{i,j}, \quad \tau_{j+N}=c_j\tau_j,
\end{equation}
for some constants $c_j$ ($N$-reduction). Correspondingly the Lax formalism is rewritten
in terms of $N\times N$ matrices with a spectral parameter $z$ as follows:
\begin{equation}\label{laxBN}
T_k \Psi=B_k \Psi, \quad B_k=U_k+t_k \Lambda,
\end{equation}
where $\Psi=(\psi_i)_{1 \leq i \leq N}$,
$U_i={\rm diag}(u_{i,1},\ldots,u_{i,N})$ and
$\displaystyle \Lambda=\sum_{i=1}^{N-1} E_{i,i+1}+z E_{N,1}$.

We next impose the similarity condition. Consider the action of the
Euler operator $T_{1,\ldots,M}$:
\begin{equation}
T_{1,\ldots,M} \Psi=B_{1,\ldots,M}\Psi,
\end{equation}
where
\begin{equation}
B_{1,\ldots,M}=T_{2,\ldots,M}(B_1)T_{3,\ldots,M}(B_2)
\cdots T_M(B_{M-1})B_{M}.
\end{equation}
Setting
\begin{equation}
A=K^{-1}B_{1,\ldots,M}, \quad K={\rm diag}(q^{N-1},q^{N-2},\ldots,1),
\end{equation}
we require that $\Psi$ should obey the similarity condition,
\begin{equation}\label{laxA}
T_{q^{N},z} \Psi=A \Psi,
\end{equation}
where $ T_{q^N,z}$ is the $q$-shift operator with respect to $z$ such
that $ T_{q^N,z}(z)=q^N z$.
The compatibility condition 
\begin{equation}\label{ABcomp}
T_{q^N,z}(B_k)A=T_k(A)B_k
\end{equation}
of (\ref{laxA}) and (\ref{laxBN}) implies in fact the homogeneity of $u_{i,j}$:
\begin{equation}
T_{1,\ldots,M} (u_{i,j})=u_{i,j}.
\end{equation}
Consistently, we may assume
\begin{equation}
T_{1,\ldots,M} (\tau_j)=\gamma_j \tau_j,
\end{equation}
for some constants $\gamma_j$.

The $N$-reduced $q$-KP hierarchy with similarity condition
gives a system of $q$-difference equations of the form
\begin{equation}\label{MNsys}
T_k(u_{i,j})=F_{k,i,j}(u), \quad (k,i=1,\ldots,M, j=1,\ldots,N),
\end{equation}
which we call $q$-Painlev\'e system of type $(M,N)$.  In (\ref{MNsys})
$F_{k,i,j}(u)$ are in general complicated rational functions of the
variables $u_{i,j}$ and $t_i$ ($i=1,\ldots,M$, $j=1,\ldots,N$).  Their
explicit form will be given in the next section.

\section{The Weyl group realization}
The time evolution and symmetry of the $q$-Painlev\'e system can be
described in the framework of a birational action of the affine Weyl
group of type $A^{(1)}_{M-1}\times A^{(1)}_{N-1}$.  To this end, we
first discuss a realization of the affine Weyl group as a group of
automorphisms of a field of rational functions, apart from the context
of the $q$-KP hierarchy.

We introduce a new set of variables $x_{i,j}$
($i=1,\ldots,M,\ j=1,\ldots,N$) and define the action of the affine Weyl
group on the field of rational functions $K=\BC(x)$ in $MN$ variables
$x_{i,j}$.  We extend the indices $i,j$ of $x_{i,j}$ to $i,j \in \BZ$ by
the condition $x_{i+M,j}=q x_{i,j}$ and $x_{i,j+N}=p x_{i,j}$ with some
fixed constants $q$ and $p$.

Define algebra automorphisms 
$\pi$, $\rho$, $r_i$ and $s_j$ ($i \in \BZ/M \BZ$,
$j \in \BZ/N \BZ$) on the field $K$
as follows:
\begin{equation}\label{rs-transformation}
\begin{array}{ll}
\pi(x_{i,j})=x_{i+1,j}, & \rho(x_{i,j})=x_{i,j+1}, \\[3mm]
r_i(x_{i,j})=p x_{i+1,j} \dfrac{P_{i,j-1}}{P_{i,j}}, &
r_i(x_{i+1,j})=p^{-1} x_{i,j} \dfrac{P_{i,j}}{P_{i,j-1}}, \\[3mm]
r_k(x_{i,j})=x_{i,j}, \quad (k \neq i,i+1),\\[3mm]
s_j(x_{i,j})=q x_{i,j+1} \dfrac{Q_{i-1,j}}{Q_{i,j}}, &
s_j(x_{i,j+1})=q^{-1} x_{i,j} \dfrac{Q_{i,j}}{Q_{i-1,j}}, \\[3mm]
s_k(x_{i,j})=x_{i,j}, \quad (k \neq j,j+1),
\end{array}
\end{equation}
where 
\begin{equation}
\begin{array}l
\displaystyle
P_{i,j}=\sum_{a=1}^{N} \left( 
\prod_{k=1}^{a-1} x_{i,j+k} 
\prod_{k=a+1}^{N} x_{i+1,j+k} \right), \\[2mm]
\displaystyle
Q_{i,j}=\sum_{a=1}^{M} \left( 
\prod_{k=1}^{a-1} x_{i+k,j} 
\prod_{k=a+1}^{M} x_{i+k,j+1} \right).
\end{array}
\end{equation}

\begin{thm} \label{AmAn}
The automorphisms $\langle \pi, r_0,r_1,\cdots,r_{M-1} \rangle$ and
$\langle \rho, s_0,s_1,\cdots,s_{N-1} \rangle$ generate the extended
affine Weyl group $\widetilde{W}(A^{(1)}_{M-1})$ and
$\widetilde{W}(A^{(1)}_{N-1})$, respectively.  Moreover these two
actions $\widetilde{W}(A^{(1)}_{M-1})$ and
$\widetilde{W}(A^{(1)}_{N-1})$ mutually commute.
\end{thm}
This theorem is proved for the special case of $p=q=1$ in \cite{KNY2}
(see also \cite{K}).  We omit the proof of Theorem \ref{AmAn} since
essentially the same argument applies to the case of general $p$ and $q$.

Let $\Gamma_i$ ($i=1,\ldots,M$)
be the translations in $\widetilde{W}(A^{(1)}_{M-1})$ defined by
\begin{equation}
\Gamma_i=r_{i-1}r_{i-2}\cdots r_{1} \pi r_{M-1}\cdots r_{i+1}r_{i}.
\end{equation}
Then $\Gamma_i$ define a commuting family of discrete flows, for which
$\widetilde{W}(A^{(1)}_{N-1})$ acts as B\"acklund transformations.
By a computation similar to that given in \cite{Y}, 
explicit formulas for the actions of $\Gamma_i$ are obtained as follows:
\begin{equation}
\Gamma_i(x_{k,j})=
\left\{
\begin{array}{ll}
p^{M-1}q x_{k,j}\dfrac{P^{k+1,\ldots,M+i}_{j-1}}{P^{k+1,\ldots,M+i}_j}
&(k=i),\\[4mm]
p^{-1} x_{k,j}\dfrac{P^{k,M+i}_{j}}{P^{k,M+i}_{j-1}}
&(k=M+i-1),\\[4mm]
p^{-1}x_{k,j}\dfrac
{P^{k+1,\ldots,M+i}_{j-1}P^{k,\ldots,M+i}_{j}}
{P^{k+1,\ldots,M+i}_{j}P^{k,\ldots,M+i}_{j-1}} &({\rm otherwise}).
\end{array}
\right.
\end{equation}
Here
\begin{equation}
P^{i,\ldots,i+r}_j=\sum_{a}
\left(\prod_{k=1}^{a_1-1}x_{i,j+k}\right)
\left(\prod_{k=a_1+1}^{a_2-1}x_{i+1,j+i}\right) \cdots
\left(\prod_{k=a_r+1}^{r N}x_{i+r,j+i}\right),
\end{equation}
and the sum $\sum_{a}$ is taken over integers $a_i$ such that $0 \leq
a_1<a_2<\ldots <a_r \leq r N$ and $1 \leq a_i-a_{i-1} \leq N$. As is
shown in \cite{KNY2}, the system for $(M,N)=(2,N)$ for $p=1$ is
equivalent to (\ref{qPAn}).

We now give an explicit correspondence between the above family of
discrete flows and the $q$-Painlev\'e systems described in the previous
section.
\begin{thm} \label{thm:main}
With the notation of 
the $q$-Painlev\'e system of type $(M,N)$ as in Section 3, 
define the variables $x_{i,j}$ by 
\begin{equation}
x_{i,j}=\frac{1}{t_i}T_{i+1,\ldots,M}(u_{i,j})
\quad (i=1,\ldots,M,\ j=1,\ldots,N)\label{ident}.
\end{equation}
Then, in terms of the $x$ variables, 
the time evolutions of the $q$-Painlev\'e system coincide
with the commuting flows defined by 
$\Gamma_k^{-1}$ $(k=1,\ldots,M)$  
under the birational action of 
$\widetilde{W}(A^{(1)}_M)=\langle r_0,\ldots,r_{M-1},\pi\rangle$
with $p=1$. Namely, we have
\begin{equation}
T_k(x_{i,j})=\Gamma_{k}^{-1}(x_{i,j})\qquad(i=1,\ldots,M,\ j=1,\ldots,N).
\end{equation}
\end{thm}

In order to prove Theorem \ref{thm:main}, the following lemma which was 
originally given in \cite{KNY2} plays a crucial role.
\begin{lem}\label{lem:aman}(Lemma 3.1 of \cite{KNY2})
For given $MN$ variables $x_{i,j}$ and $MN$ unknowns $x'_{i,j}$ 
$(i=1,\ldots,M$, $j=1,\ldots,N)$, 
we denote $X_i={\rm diag}(x_{i,1},\ldots,x_{i,N})$ and
$X'_i={\rm diag}(x'_{i,1},\ldots,x'_{i,N})$.
Then the system of algebraic equations
\begin{equation}
 (X_1+\Lambda)\cdots(X_M+\Lambda)=
 (X'_1+\Lambda)\cdots(X'_M+\Lambda),
\end{equation}
has $M!$ solutions.  Each of the solution corresponds to a permutation
$\sigma \in S_M$ and characterized by additional conditions
\begin{equation}
 x'_{i,1}x'_{i,2}\cdots x'_{i,N}=
x_{\sigma(i),1}x_{\sigma(i),2}\cdots x_{\sigma(i),N}
\quad (i=1,\ldots,M).
\end{equation}
Moreover, if $\sigma$  is given as a product
$\sigma=\sigma_{i_1}\cdots \sigma_{i_k}$, where $\sigma_i=(i,i+1)$,
then the corresponding solution is explicitly expressed as
\begin{equation}
 x'_{i,j}=r_{i_1}\cdots r_{i_k}(x_{i,j}),
\end{equation}
by means of the birational Weyl group action of 
$W(A^{(1)}_{M-1})=\langle r_1, \ldots, r_{M-1} \rangle$ with $p=1$
$(\ref{rs-transformation})$.
\end{lem}

\noindent\textit{Proof of Theorem \ref{thm:main}.}\hskip2mm
We first remark that, for each $i=1,\ldots,M$, 
the product $u_{i,1}\cdots u_{i,N}$ is invariant 
with respect to the $q$-shift operators $T_k$ ($k=1,\ldots,M$). 
Hence, we have
\begin{equation}
x_{i,1}\cdots x_{i,N}=t_i^{-N}
T_{i+1,\ldots,M}(u_{i,1}\cdots u_{i,N}) = 
t_i^{-N}u_{i,1}\cdots u_{i,N}
\end{equation}
and 
\begin{equation}
T_k(x_{i,1}\ldots x_{i,N})=q^{-N\delta_{i,k}} x_{i,1}\cdots x_{i,N}.
\end{equation}
We now prove that, for each $k=1,\ldots,M$, 
the action of $T_k^{-1}$ on the $x$ variables 
coincides with that of 
$\Gamma_k=r_{k-1}\cdots r_1\pi r_{M-1}\cdots r_k$
(with $p=1$).  
Since $T_{1,\ldots,M}=T_k T_{M}T_{M-1}\cdots\widehat{T}_k\ldots T_{1}$, 
the matrix $B_{1,\ldots,M}$ is expressed alternatively as
\begin{eqnarray}
&&B_{1,\ldots,M}=T_{2,\ldots,M}(B_1)T_{3,\ldots,M}(B_2)
\cdots T_{k,\ldots,M}(B_{k-1})\nonumber\\
&&\phantom{B_{1,\ldots,M}=}
\cdot\ T_k\big[T_{k+2,\ldots,M}(B_{k+1})
\cdots T_{M}(B_{M-1}) B_M] B_k. 
\end{eqnarray}
This implies 
\begin{eqnarray}
&&
(X_1+\Lambda)\cdots(X_M+\Lambda)
=(X_1+\Lambda)\cdots(X_{k-1}+\Lambda)
\nonumber\\
&&
\phantom{(X_1+\Lambda)\cdots(X_M+\Lambda)}
\cdot \,T_k\big[(X_{k+1}+\Lambda)
\cdots(X_M+\Lambda)\big](t_k^{-1}U_k+\Lambda).
\end{eqnarray}
In the right-hand side, the products of diagonal entries 
of individual factors are arranged as 
\begin{equation}
\begin{array}{ll}\smallskip
x_{i,1}\cdots x_{i,N}\quad & (i=1,\ldots,k-1),
\cr\smallskip
T_k(x_{i,1}\cdots x_{i,N})=x_{i,1}\cdots x_{i_N}
& (i=k+1,\ldots,N),\cr
t_k^{-N} u_{k,1}\cdots u_{k,N}=x_{k,1}\cdots x_{k,N}.
\end{array}
\end{equation}
according to the cyclic permutation $(k,k+1,\ldots,M)$.
Hence by Lemma \ref{AmAn}, we obtain
\begin{equation}\label{eq:rr1}
\begin{array}{rl}\smallskip
x_{i,j}=r_{k}\cdots r_{M-1}(x_{i,j})\quad &(i=1,\ldots,k-1), 
\cr\smallskip
T_k(x_{i+1,j})=r_{k}\cdots r_{M-1}(x_{i,j})\quad&(i=k,\ldots,M-1),\cr
t_k^{-1} u_{k,j}=r_{k}\cdots r_{M-1}(x_{M,j}),\ \ 
\end{array} 
\end{equation} 
for all $j=1,\ldots,N$. 
Similarly, from 
$T_{1,\ldots,M}=T_{M}\cdots \widehat{T}_k \cdots T_1 T_k$, 
we obtain 
\begin{eqnarray}
&&B_{1,\ldots,M}= 
T_k^{-1}\big[
T_{1,\ldots,M}(B_k)
T_{2,\ldots,M}(B_1)T_{3,\ldots,M}(B_2)
\cdots T_{k,\ldots,M}(B_{k-1})\big]\nonumber\\
&&\phantom{B_{1,\ldots,M}=}
\cdot T_{k+2,\ldots,M}(B_{k+1})
\cdots T_{M}(B_{M-1}) B_M,
\end{eqnarray}
hence
\begin{eqnarray}
&&
(X_1+\Lambda)\cdots(X_M+\Lambda)
=
(t_k^{-1}T_k^{-1}(U_k)+\Lambda)
T_k^{-1}\big[
(X_1+\Lambda)\cdots(X_{k-1}+\Lambda)\big]
\nonumber\\
&&
\phantom{(X_1+\Lambda)\cdots(X_M+\Lambda)=}
\cdot (X_{k+1}+\Lambda)
\cdots(X_M+\Lambda). 
\end{eqnarray}
This time the factorization on 
the right-hand side corresponds to the cyclic 
permutation $(k,k-1,\ldots,1)$; hence we have
\begin{equation}\label{eq:rr2}
\begin{array}{rl}\smallskip
t_k^{-1}T_k^{-1}(u_{k,j})=r_{k-1}\cdots r_{1}(x_{1,j}),\quad &
\cr\smallskip
T_k^{-1}(x_{i-1,j})=r_{k-1}\cdots r_{1}(x_{i,j})\quad&(i=2,\ldots,k),\cr
x_{i,j}=r_{k-1}\cdots r_{1}(x_{i,j})\quad&(i=k+1,\ldots,M)
\end{array} 
\end{equation} 
for all $j=1,\ldots,N$. 
By using (\ref{eq:rr1}) and (\ref{eq:rr2}), we can determine
the action of $T_k^{-1}$ as follows:
\begin{equation}
\begin{array}{ll}
\smallskip
T_k^{-1} r_{k}\cdots r_{M-1}(x_{i,j})=T_k^{-1}(x_{i,j})=
r_{k-1}\cdots r_1(x_{i+1,j})\ \ &(i=1,\ldots,k-1),
\cr\smallskip
T_k^{-1} r_{k}\cdots r_{M-1}(x_{i,j})=x_{i+1,j}
=r_{k-1}\cdots r_1(x_{i+1,j})\ \ &(i=k,\ldots,M-1),\cr
T_k^{-1} r_{k}\cdots{x_{M,j}}=
qt_k^{-1}T_{k}^{-1}(u_{k,j})=r_{k-1}\cdots r_1(q x_{1,j}),
\end{array}
\end{equation}
for all $j=1,\ldots,N$.  Namely we have
\begin{equation}
T_k^{-1}r_{k}\cdots r_{M-1}(x_{i,j})=r_{k-1}\cdots r_1 \pi(x_{i,j})
\quad(i=1,\ldots,M,\,j=1,\ldots,N). 
\end{equation}
This means that the action of $T_k^{-1}$ on the $x$ variables 
coincides with that of 
$\Gamma_k=r_{k-1}\cdots r_1 \pi r_{M-1}\cdots r_k$. 
\qed

\medskip

We now construct the rational solutions for the $q$-Painlev\'e system.

We say that a sequence of partitions $\lambda(i)$ ($i \in \BZ/N \BZ$) is
an $N$-reduced chain if $\lambda(i)\equiv(k_i,\lambda(i-1))$ for some
$k_i \in \BZ$. For instance $\lambda(0)=(1,1)$, $\lambda(1)=(2,1,1)$ and
$\lambda(2)=(2,2,1,1)$ is an example of 3-reduced chain, where $k_1=2$,
$k_2=2$, $k_3=k_0=-4$.

For any $N$-reduced chain
$\lambda(i)=(\lambda_1(i),\lambda_2(i),\ldots)$ $(i\in\BZ/N\BZ)$ given,
we put $\tau_i=S_{\lambda(i)}(t_1,\ldots,t_M)$.   Then from Proposition
\ref{qschur-sol}, the $\tau$ functions $\tau_i$ ($i \in \BZ/N\BZ$) give
a similarity solution of the $N$-reduced $q$-KP hierarchy such that
$T_{1,\ldots,M}(\tau_i)=\gamma_i\tau_i$ with
$\gamma_i=|\lambda(i)|=\lambda_1(i)+\lambda_2(i)+\cdots$.

\begin{cor} For any $N$-reduced chain of $\tau$ functions $\tau_i$
 defined as above, we put
\begin{equation}\label{xijsol}
x_{i,j}=\dfrac{1}{t_i}\dfrac{T_{i,\ldots,M}(\tau_j)T_{i+1,\ldots,M}(\tau_{j-1})}
{T_{i+1,\ldots,M}(\tau_j)T_{i,\ldots,M}(\tau_{j-1})}\quad (i=1,\ldots, M,\ j=1,\ldots,N).
\end{equation}
Then $x_{i,j}$ solve the $q$-Painlev\'e system of type $(M,N)$
\begin{equation}
x_{i,j}(t_1,\ldots,q^{-1} t_k,\ldots,t_M)=\Gamma_k(x_{i,j})\quad (k=1,\ldots, M).
\end{equation}
\end{cor}


We finally remark on the Toda type bilinear $q$-difference equations
satisfied by $q$-Schur functions. For simplicity,
we consider the case of $M=2$, namely the case with two time variables $t=(t_1,t_2)$.
For a $q$-Schur function $\tau=S_{\lambda}(t)$ let us put
$\tau_{k,\cdots,l}=S_{(l,\cdots,k,\lambda)}(t)$. 
It follows from Proposition \ref{qS-Back} that 
\begin{equation}\label{tau-Back12}
\begin{array}l
t_1 \tau_{k,l} T_1(\tau)+\tau_{k}T_1(\tau_l)-\tau_l T_1(\tau_k)=0,\\[3mm]
t_2 \tau_{k,l} T_2(\tau)+\tau_{k}T_2(\tau_l)-\tau_l T_2(\tau_k)=0.
\end{array}
\end{equation}
Applying $T_1$ on the second equation, and using the similarity
condition, we have
\begin{equation}\label{tau-Back13}
t_2 T_1(\tau_{k,l}) \tau+T_1(\tau_{k})q^{l}\tau_l
-T_1(\tau_l) q^{k}\tau_k=0.
\end{equation}
Combining (\ref{tau-Back13}) and the first equation of
(\ref{tau-Back12}), we obtain
\begin{equation}\label{qToda}
q^k t_1 T_1(\tau)\tau_{kl}+t_2 T_1(\tau_{kl})\tau=
(q^k-q^l) T_1(\tau_k)\tau_l.
\end{equation}
Define monic polynomial $Q_{\lambda}(x)$
as $Q_{\lambda}(x)=c_{\lambda}^{-1} S_{\lambda}(x,1)$.
Here the normalization constants are given, in terms of
the hook-length \cite{Mac} $h_{i,j}=\lambda_i-j+\lambda'_j-i+1$ of the 
partition $\lambda$, as 
\begin{equation}
c_{\lambda}=q^{\sum_i (i-1)\lambda_i}
\prod_{(i,j) \in \lambda}(1-q^{h_{i,j}})^{-1}.
\end{equation}
Then (\ref{qToda}) yields
\begin{equation}\label{qToda2}
Q_{(k,\lambda)}(qx)Q_{(l,\lambda)}(x)=
Q_{(l-1,k,\lambda)}(qx)Q_{\lambda}(x)+
x q^k Q_{\lambda}(qx)Q_{(l-1,k,\lambda)}(x).
\end{equation}
In the 3-reduced case, (\ref{qToda2}) coincides with the Toda equations
for the $q$-Okamoto polynomials(\cite{KNY2} Theorem 2.7).  At the same
time, from the determinant formula (\ref{qSchur}), we obtain the
Jacobi-Trudi type formula for the $q$-Okamoto polynomials; in this
particular case, the entries are essentially the continuous $q$-Hermite
polynomials.

\end{document}